\documentclass[fleqn,usenatbib]{mnras}

\usepackage{newtxtext,newtxmath}
\usepackage[T1]{fontenc}

\DeclareRobustCommand{\VAN}[3]{#2}
\let\VANthebibliography\thebibliography
\def\thebibliography{\DeclareRobustCommand{\VAN}[3]{##3}\VANthebibliography}

\usepackage{graphicx}	
\usepackage{xcolor}

\usepackage{amsmath}
\graphicspath{{./}{figures/}}

\def\kmsMpc{\ensuremath{\text{km} \, \text{s}^{-1} \text{Mpc}^{-1}}}

\title[$H_0$ estimation with dark standard sirens and galaxy cluster catalogues]{A Hubble constant estimation with dark standard sirens and galaxy cluster catalogues}

\author[F.~Beirnaert et al.]{
Freija Beirnaert,\textsuperscript{1}\thanks{E-mail: freija.beirnaert@ugent.be (FB)},
Gergely Dálya\textsuperscript{2,3}
and
Archisman Ghosh\textsuperscript{1} \\
\textsuperscript{1}{Department of Physics \& Astronomy, Ghent University, Proeftuinstraat 86, 9000 Ghent, Belgium} \\
\textsuperscript{2}{L2IT, Laboratoire des 2 Infinis - Toulouse, Université de Toulouse, CNRS/IN2P3, UPS, F-31062 Toulouse Cedex 9, France
} \\
\textsuperscript{3}{MTA-ELTE Astrophysics Research Group, 1117 Budapest, Hungary} \\
}

\date{Accepted XXX. Received YYY; in original form ZZZ}

\pubyear{2025}

\begin{document}
\label{firstpage}
\pagerange{\pageref{firstpage}--\pageref{lastpage}}
\maketitle

\begin{abstract}
In this paper, we explore the possibility of using galaxy cluster catalogues to provide redshift support for a gravitational-wave dark standard siren measurement of the Hubble constant $H_0$. We adapt the cosmology inference pipeline \texttt{gwcosmo} to handle galaxy cluster catalogues. Together with binary black holes from the GWTC-3, we use galaxy cluster data from the PSZ2 and the eRASS catalogues. With these catalogues, we obtain $H_0 = 77^{+10}_{-10}$ and $81^{+8}_{-8} \, \kmsMpc$ respectively, which demonstrates improvements on precision by factors of $10\%$ and $38\%$ respectively over the traditional galaxy catalogue result. This exploratory work paves the way towards precise and accurate cosmography making use of distant compact binary mergers from upcoming observing runs of the LIGO-Virgo-KAGRA detector network and future gravitational-wave observatories.
\end{abstract}


\maketitle

\section{Introduction}
\label{introduction}

Merging compact binaries observed in gravitational waves (GWs) are standard distance indicators or standard sirens~\citep{Schutz:1986gp,Holz:2005df}. With complementary information on the source redshift, they can be used to measure cosmological parameters such as the Hubble constant $H_0$. Since the multimessenger observation of GW170817, a new avenue has opened up in cosmology, making use of the observed GW sources to measure late-time cosmological parameters, including $H_0$ and other parameters of $\Lambda$CDM cosmology~\citep{LIGOScientific:2017adf,DES:2019ccw,LIGOScientific:2019zcs,LIGOScientific:2021aug,Palmese:2021mjm,Bom:2024afj}, as well as parameters of modified gravity~\citep{Finke:2021aom,Chen:2023wpj}. Apart from GW170817, none of the compact binaries observed until now were accompanied by bright electromagnetic (EM) counterparts. In absence of bright counterparts, the complementary redshift information can still come from the observed mass distribution of the GW sources for {\em spectral sirens}~\citep{Farr:2019twy,Mastrogiovanni:2021wsd,Mukherjee:2021rtw,Karathanasis:2022rtr,Ezquiaga:2022zkx,Farah:2024xub,2024arXiv240402522M,2024arXiv241023541N}, or statistically from the redshifts of the potential host galaxies in a galaxy catalogue for {\em dark standard sirens}~\citep{LIGOScientific:2018gmd,DES:2019ccw,Gray:2019ksv,LIGOScientific:2019zcs,Gair:2022zsa}. Other ways of obtaining the redshift, such as via the internal physics of neutron stars~\cite{Messenger:2011gi,Chatterjee:2021xrm}, or the spatial clustering scale of GW localizations~\citep{Oguri:2016dgk,Mukherjee:2020hyn,Ghosh:2023ksl} are promising for future observatories. A detailed overview of the different methodologies and the prospects are available in \cite{Chen:2024gdn,2025arXiv250200239P}.

The most up-to-date $H_0$ measurement from LIGO-Virgo-KAGRA (LVK)~\citep{LIGOScientific:2021aug} employs events from the Third Gravitational-Wave Transient Catalogue (GWTC-3) \citep{KAGRA:2021vkt} and galaxy data from the GLADE+ catalogue \citep{Dalya:2021ewn}. There are efforts to extend the GLADE+ catalogue to higher redshifts and with better photometric data. The upgraded version, UpGLADE is expected to be complete up to $z \simeq 0.25$ in the $g$-band for $50\%$ of its sky coverage, in terms of containing the host galaxy of GW events. In parallel, there are attempts to make use of other galaxy catalogues including the Dark Energy Survey and the Dark Energy Spectroscopic Instrument Legacy Survey. However in spite of these significant enhancements, GW events at luminosity distances greater than $d_L\gtrsim1\,\text{Gpc}$ (corresponding to $z\gtrsim0.25$) are expected to be left without any significant support from galaxy catalogues.

The reach of the method can be extended by using {\em galaxy cluster catalogues}, which are complete up to a higher redshift and of comparable range to current GW detections. This makes high-$d_L$ events potentially more informative. By making more use of these distant events, we also enter the non-linear regime of cosmology, eventually enabling estimation of parameters beyond $H_0$. 
In the traditional method using galaxy catalogues, one weights each host galaxy with a probability in proportion to its luminosity. The assumption is that galaxy luminosity is a measure of the total stellar mass, and traces the total matter distribution, which is also traced by the GW sources. While a galaxy's host probability is derived from its luminosity, galaxy cluster catalogues often entail a more direct measurement of the galaxy cluster mass. One can thus construct a mass-weighted redshift prior which potentially provides a more direct link to the underlying matter distribution.

Not all GW events are necessarily hosted in the centres of galaxy clusters. In fact only about $15\%$ of the total mass of the Universe (including matter and dark matter) resides in the central nodes of the cosmic web~\citep{Hellwing:2020fzr}. It is worth highlighting that the idea behind using galaxy cluster catalogues is {\em not} that all GW events are hosted in the centres of galaxy clusters, but rather that the galaxy clusters serve as tracers of the underlying (stellar) mass, which in turn is expected to be proportional to the GW merger rate density in the source frame. The nodes are expected to {\em trace} the filaments and the walls of the cosmic web where many of the host galaxies may be located. Typical offsets between host galaxy redshifts and cluster redshifts are much smaller compared to the typical GW distance uncertainties (which are $\mathcal{O}(10\%)$ for current detectors). Thus one does not expect a large systematic bias by replacing the redshift of a galaxy with that of its closest cluster.  Baryonic matter in voids left out of the above computation largely consists of hot and cold diffuse gas~\citep{Haider:2015caa}, with little to no chance of hosting GW mergers.

Galaxy clusters are biased tracers of the underlying mass distribution~\citep{1998astro.ph.10088P,Seppi:2024cfg}. The corresponding bias parameter differs from the bias associated with GW sources, which traces the spatial distribution of the GW event rate ~\citep{Oguri:2016dgk}. The connection between these two bias parameters is governed by complex astrophysical processes that are not fully understood. Although future GW observations are expected to constrain the GW bias with respect to the galaxy cluster bias parameter, we currently assume that both are equal for simplicity.

In this paper, we explore the use of galaxy cluster catalogues with the cosmology inference pipeline \texttt{gwcosmo}~\citep{Gray:2019ksv,Gray:2021sew,Gray:2023wgj}. In particular, we use the Second Planck Catalogue of Sunyaev-Zel'dovich Sources (PSZ2)~\citep{Planck:2015koh} and the eROSITA All Sky Survey (eRASS) X-ray catalogue~\citep{Bulbul:2024mfj} together with GW events from the GWTC-3 catalogue~\citep{KAGRA:2021vkt}. We observe a significant contribution from the EM sector to the dark standard siren result beyond the ``spectral'' contribution obtained from the observed mass distribution alone.

The rest of this paper is organized as follows. In Sec.~\ref{sec: method} we go over the theory and implementation adapted from~\cite{Gray:2023wgj}. In Sec.~\ref{sec: data}, we give an overview of the GW and galaxy cluster catalogue data used in this analysis. The resulting $H_0$ posteriors are shown in Sec.~\ref{sec: results}. A discussion follows in Sec.~\ref{sec: discussion} where we also outline the potential systematic effects in our analysis. Finally, we draw our conclusions in Sec.~\ref{sec: conclusion} and discuss future developments.

\section{Method}
\label{sec: method}

In this paper, we adapt the most recent implementation of the \texttt{gwcosmo} code \citep{Gray:2023wgj} to infer $H_0$ using observed compact binary coalescences (CBCs) and galaxy cluster catalogues. We combine the distance information from GW observations with the redshift information constructed as a {\em ``line-of-sight redshift prior''} (LOS-$z$ prior) using the potential host galaxy cluster redshifts and masses. The use of galaxy clusters instead of galaxies leads to changes only in the LOS-$z$ prior of \cite{Gray:2023wgj}, in which the redshift information is encoded. We define the LOS-$z$ prior as a weighted combination of in-catalogue ($C$) and out-of-catalogue contributions as
\begin{equation}
    \begin{split}
        p(z |\Omega_i, \Lambda, s) = \frac{1}{p(s | \Omega_i, \Lambda)} &[ p(C|\Omega_i, \Lambda)  f_{\text{in}}(z) +f_{\text{out}}(z) ]\,,\\
    \end{split}
\end{equation}
where $f_{\text{in}}(z)$ and $f_{\text{out}}(z)$ are the in- and out-of-catalogue components.

The in-catalogue component models the galaxy cluster redshifts as Gaussian distributions centered around their best redshift estimates and weighted by their masses as
\begin{align}
f_{\text{in}}(z) = \frac{1}{N_C(\Omega_i)} \sum_{k=1}^{N_C(\Omega_i)} p(z | \hat{z}_k) \, p(M)\,,
\end{align}
where $p(z | \hat{z}_k)$ is the redshift distribution of galaxy cluster $k$, which has a width either modelled or provided by the cluster catalogue as described in Sec.~\ref{sec: cluster catalogue data}.

The out-of-catalogue component accounts for the incompleteness of the galaxy cluster catalogue due to selection effects. This is modelled by a prior on the galaxy cluster mass which can based on \textit{Press-Schechter} (PS) \citep{Press:1973iz} or \textit{Sheth-Tormen} (ST) \citep{2001MNRAS.323....1S} formalism, of which the mass functions are given by
\begin{equation}
        p_{\text{PS}}(M)\propto M^{-2}\nu\left|\frac{d \ln \sigma}{d \ln M}\right|e^{-\nu^2 / 2}
        \label{eq: PS}
\end{equation}
and
\begin{equation}
   p_{\text{ST}}(M)\propto M^{-2} \nu\left[1+\left(\frac{1}{\left(a \nu^2\right)^p}\right)\right] e^{-a \nu^2 / 2},
   \label{eq: ST}
\end{equation}
where $\nu=\frac{\delta_c}{\sigma(M)}$ with $\delta_c=1.686$ the critical overdensity for collapse and $\sigma(M)=\left(\frac{M}{M_*}\right)^{-(n+3)/6}$ the RMS density fluctuation on the scale of $M$. Further, $a=0.707$ is an ellipticity fudge factor and $p=0.3$ an empirical parameter. 

In the local universe, we can employ the following parametrization based on optical virial mass estimates from \cite{Girardi:1998rm}.
\begin{equation}
    p(M)\,=\,n^*\,\left({M\over M^*}\right)^{-1}e^{-M/M^*}
    \label{eq: Girardi}
\end{equation}
with \mbox{$n^*=2.6^{+0.5}_{-0.4}\times 10^{-5}\,(h^{-1}
\text{Mpc})^{-3}(10^{14}h^{-1}M_\odot)^{-1}$} and \\[0.2em]
\mbox{$M^*=2.6^{+0.8}_{-0.6} \times
10^{14} \, h^{-1}M_\odot$}. \\[-0.8em]

This function is integrated in the regions where we assume the catalogue to be empty. For low redshifts, the reach of the mass range is modelled by the galaxy cluster mass selection function $M(z)$, below which no clusters are detected. In the region beyond $z_{\text{cut}}$ where we assume the catalogue to be empty, mass prior constitutes the full LOS-$z$ prior, such that the out-of-catalogue contribution becomes
\begin{equation}
    \begin{split}
        f_{\text{out}}(z) = & \, \Theta(z_\text{cut}-z) \int_{M_\text{min}}^{M(z)} p(z, M | \Lambda) \, p(s | z, M) \, dM \\
    & + \Theta(z-z_\text{cut}) \int_{M_\text{min}}^{M_\text{max}} p(z, M | \Lambda) \, p(s | z, M) \, dM.
    \end{split}
\end{equation}

The prior on the redshift is taken to be proportional to the differential comoving volume, $p(z)\propto dV_c(z)/dz$. The sky location prior $p(\Omega)$ is taken to be uniform in the sky. Finally, we choose a prior on $H_0$ uniform $\in [20,140]$ \kmsMpc. 
The cosmology, compact binary mass distribution and rate evolution models are described in further detail in App.~\ref{app: models}. For this study, we have fixed the CBC population parameters. The formalism in \cite{Gray:2023wgj} already allows for relaxing these assumptions towards a more realistic analysis in the future.

\section{Data}
\label{sec: data}
\subsection{GW events}

Our analysis is carried out with binary black hole (BBH) events from GWTC-3 \citep{LIGOScientific:2019lzm} with signal-to-noise ratio (SNR) values above an SNR threshold 9. Our motivation for choosing a wider selection of GW events than in \cite{KAGRA:2023pio} is that galaxy cluster catalogues have significant support out to higher redshifts, and from corresponding large distances, from where we typically expect low-SNR events. The selection of events which we finally use is shown in Fig.~\ref{fig: event h0 combined fixed populations} and their GW inference parameters can be found in \cite{LIGOScientific:2019lzm}; we use the publicly available GW parameter estimation posterior samples for these events.\footnote{\url{https://gwosc.org/GWTC-3/}} The GW selection function of \cite{Gray:2023wgj} is adjusted appropriately to account for the lower SNR threshold.

\subsection{Galaxy cluster catalogues} \label{sec: cluster catalogue data}
In this work, we employ two types of galaxy cluster catalogues, the PSZ2 and the eRASS catalogues, with sky coverages of 83.6\% and 31.8\% respectively.

\begin{figure}
    \centering
    \includegraphics[width=0.47\textwidth]{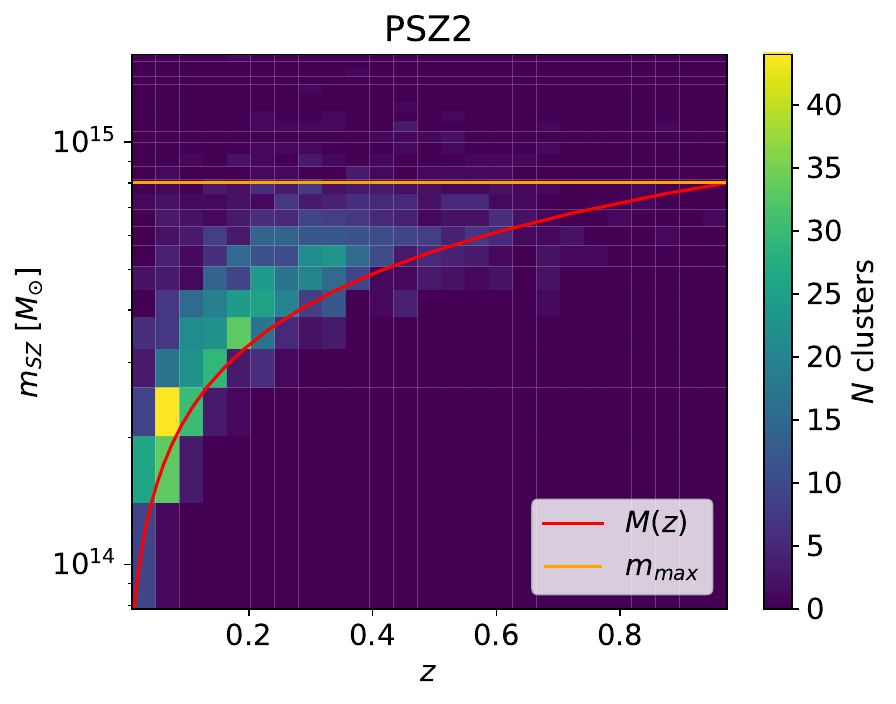}
    \includegraphics[width=0.47\textwidth]{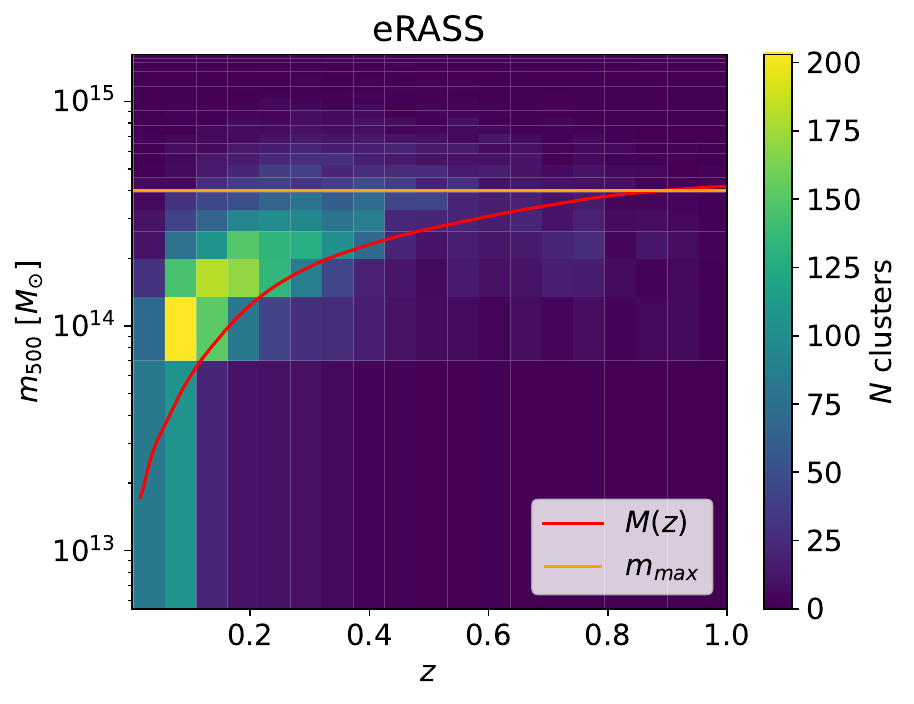}
    \caption{Mass and redshift data for the galaxy cluster catalogues PSZ2 (top panel) and eRASS (bottom panel), along with the used lower and upper mass limits used in the analysis.}
    \label{fig: z m hists}
\end{figure}

The PSZ2 galaxy cluster catalogue\footnote{\url{https://cdsarc.cds.unistra.fr/viz-bin/Cat?J/A+A/594/A27}} \citep{Planck:2015koh} from the Planck Collaboration has been created using the Sunyaev-Zel'dovich (SZ) effect, by which galaxy cluster candidates are discovered via the inverse Compton scattering of the Cosmic Microwave Background (CMB), leading to a local spectral distortion from which the mass is derived. The galaxy cluster candidates are subsequently cross-matched with galaxy clusters from optical surveys which provide a redshift estimate. As the catalogue does not provide redshift uncertainties, we model these within the general cluster size range between 1 and 10 Mpc, proportional to the cluster mass. Although in principle the SZ effect has a constant lower mass limit, the data in Fig.~\ref{fig: z m hists} (top panel) display a rise in galaxy cluster mass for increasing redshift, such that the selection function can be approximated by a power law.

The eRASS galaxy cluster catalogue\footnote{\url{https://cdsarc.cds.unistra.fr/viz-bin/cat/J/A+A/685/A106}} \citep{Bulbul:2024mfj} from the SRG/eROSITA All-Sky Survey is a collection of galaxy cluster candidates identified through X-ray observations of the diffuse thermal emission from the hot intracluster medium (ICM). These spatially extended objects are distinguished from point-like sources such as active galactic nuclei (AGN) through a combination of source detection algorithms and morphological classification. Their properties such as mass and redshift are inferred (both with 1-$\sigma$ uncertainties) using scaling relations between X-ray observables (primarily luminosity and temperature) which are calibrated through both simulations and existing datasets of galaxy clusters. The selection function used in Fig.~\ref{fig: z m hists} (bottom panel) is taken from Fig.~7 in \cite{Bulbul:2024mfj}.

\begin{figure}
    \centering
    \includegraphics[width=0.45\textwidth]{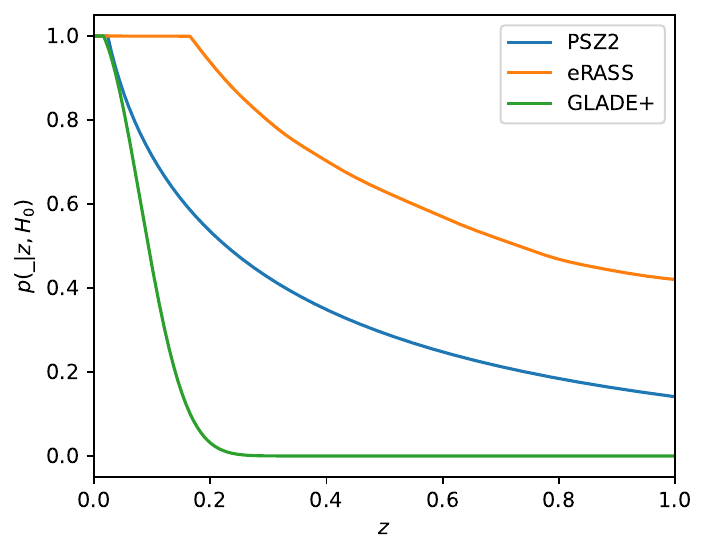}
    \caption{Completeness fraction of the PSZ2 and eRASS galaxy cluster catalogues and the GLADE+ galaxy catalogue (in the $K$-band) indicating the probability that the catalogue contains the host cluster or galaxy of a GW event, as a function of redshift for $H_0=70\,\kmsMpc$ and $\Omega_M = 0.3065$. 
    }
    \label{fig: completeness}
\end{figure}

The effective mass limits lead to the completeness curves shown in Fig.~\ref{fig: completeness}, where they are compared with the completeness of the GLADE+ catalogue in the $K$-band. This figure illustrates that both galaxy cluster catalogues which we use are significantly more complete up to higher redshifts than galaxy catalogues. GLADE+ and PSZ2 are considered to be fully complete up to a redshift of 0.02, while eRASS remains complete up to 0.16.

\section{Results}
\label{sec: results}

\begin{figure*}
    \centering
    \includegraphics[width=\textwidth]{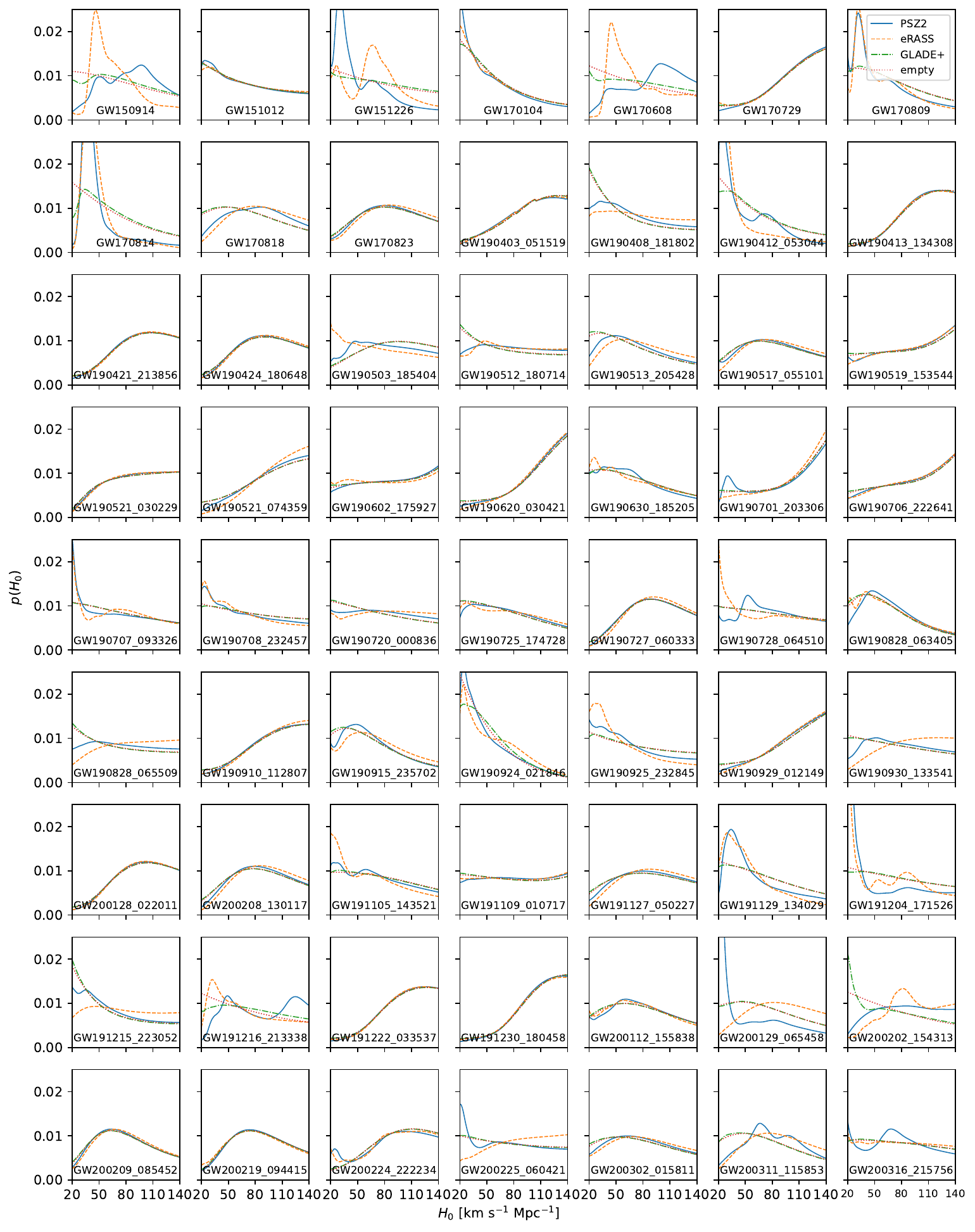}
    \caption{Posterior probability distributions on $H_0$ from individual BBH events (with SNR$>9$) from GWTC-3 and PSZ2 and eRASS galaxy cluster catalogues. The results are compared with the results using the traditional GLADE+ galaxy catalogue and the ``empty catalogue'' case. The in-catalogue contribution from galaxy cluster catalogues is evident in features such as peaks in the distributions for several of the events.}
    \label{fig: event h0 combined fixed populations}
\end{figure*}

The posterior distributions on $H_0$ from the individual events of our selection (BBHs from GWTC-3 with SNR $>$ 9) are shown in Fig.~\ref{fig: event h0 combined fixed populations}. We compare our results obtained with the PSZ2 and the eRASS catalogues with the GLADE+ and the ``empty catalogue'' results, the latter coming solely from the out-of-catalogue part via the observed mass distribution. For the first time, we see an in-catalogue contribution in the $H_0$ posterior for a large portion of the events, clearly visible as peaks in the distributions. An in-catalogue contribution from such a significant fraction of events was absent in all dark standard siren analyses carried out until now (compare e.g.~with Fig.~7 of \cite{LIGOScientific:2021aug}). In particular, only well-localized individual events such as GW170814~\citep{DES:2019ccw} and GW190814 \citep{LIGOScientific:2020zkf,DES:2020nay} were seen to contribute via galaxy catalogues.\footnote{Since we consider only BBHs, GW190814 is left out from our analysis. GW190814, being a nearby event ($d_L\sim240\,\text{Mpc}$), actually performs better with GLADE+, and shows relatively no redshift support in galaxy cluster catalogues which we consider.} In this sense, our analysis demonstrates the original idea in \cite{Schutz:1986gp} with real data: how an ensemble of GW events with redshift support from a set of identified and misidentified entities (galaxy clusters in our case) drives the $H_0$ estimate to a unimodal value. 

\begin{figure}
    \centering
    \includegraphics[width=0.47\textwidth]{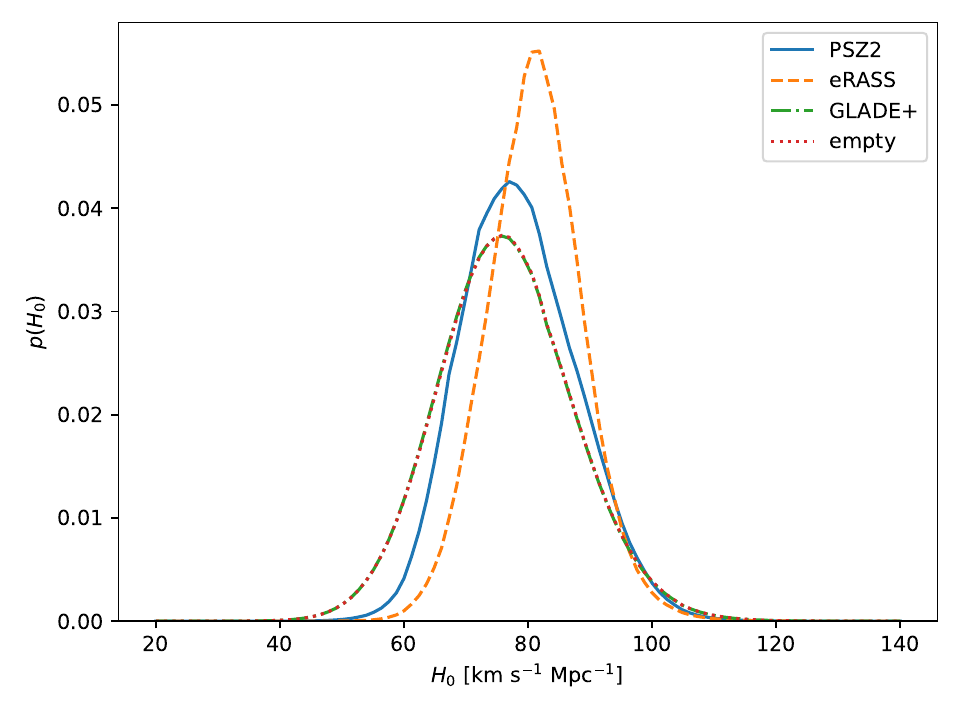}
    \caption{Combined posterior probability distribution on $H_0$ from the set of events in Fig.~\ref{fig: event h0 combined fixed populations}. The galaxy cluster catalogues PSZ2 and eRASS respectively lead to improvements on precision of $10\%$ and $38\%$ over the traditional result from the GLADE+ galaxy catalogue as well as the ``empty catalogue'' case.}
    \label{fig: h0 combined fixed populations}
\end{figure}

The combined $H_0$ posterior from our selection of GW events is shown in Fig.~\ref{fig: h0 combined fixed populations}. 
We infer $H_0$ values of $77^{+10}_{-10}$ and $81^{+8}_{-8} \, \kmsMpc$ respectively with PSZ2 and eRASS catalogues, while both the GLADE+ and the ``empty catalogue'' yield values of $76^{+11}_{-11}\, \kmsMpc$ for GWTC-3 BBHs (MAP and $68\%$ CI reported in all cases). We conclude that the $H_0$ uncertainties shrink by factors of $1.10$ and $1.38$ respectively with PSZ2 and eRASS galaxy cluster catalogues.

\section{Discussion}
\label{sec: discussion}

Galaxy cluster catalogues help us reach higher redshifts where there is little to no support in traditional galaxy catalogues. At the same time they help us overcome some of the potential systematic effects in the current method. For example, in the current method with galaxy catalogues, it is assumed that luminosity is a tracer of the underlying mass distribution, an assumption that can be brought into question~\citep{2024arXiv240507904P,Hanselman:2024hqy}. Galaxy clusters provide a more direct probe of the underlying mass distribution which can be related to the distribution of GW events with fewer assumptions. The method is however susceptible to uncertainties in the mass estimates of the galaxy clusters, and the uncertainties in parameter $M^*$ of the Press-Schechter function (which enters the out-of-catalogue part). Galaxy redshifts are often determined via photometry; for the GLADE+ catalogue, for example, precise spectroscopic redshifts are available for only $1\%$ of the galaxies. Photometric redshifts are typically obtained by assuming redshift uncertainty models and fitting measurements in different photometric bands across a large set of galaxies. Mismodelling of galaxy redshift distributions is known to bias the $H_0$ measurement downstream~\citep{Turski:2023lxq}. Galaxy clusters on the contrary have precise redshift measurements, usually via cross-matching with identified bright counterparts or obtained by averaging redshifts of member galaxies. Uncertainties arising from redshift measurements can hence be expected to be smaller for galaxy clusters than for individual galaxies. Peculiar velocities of galaxy clusters can potentially affect the measurement accuracy. Estimates from ~\cite{Bahcall:1996bc} suggest that peculiar velocities of galaxy clusters can be as high as $600\,\text{km}\,\text{s}^{-1}$. This corresponds to a  a redshift error of $\Delta z = 0.002$ corresponding to $\sim8\,\text{Mpc}$ (with $H_0=70\,\kmsMpc$). Our analysis below however demonstrates that this is not a very significant effect on the whole.

\begin{figure}
    \centering
    \includegraphics[width=0.47\textwidth]{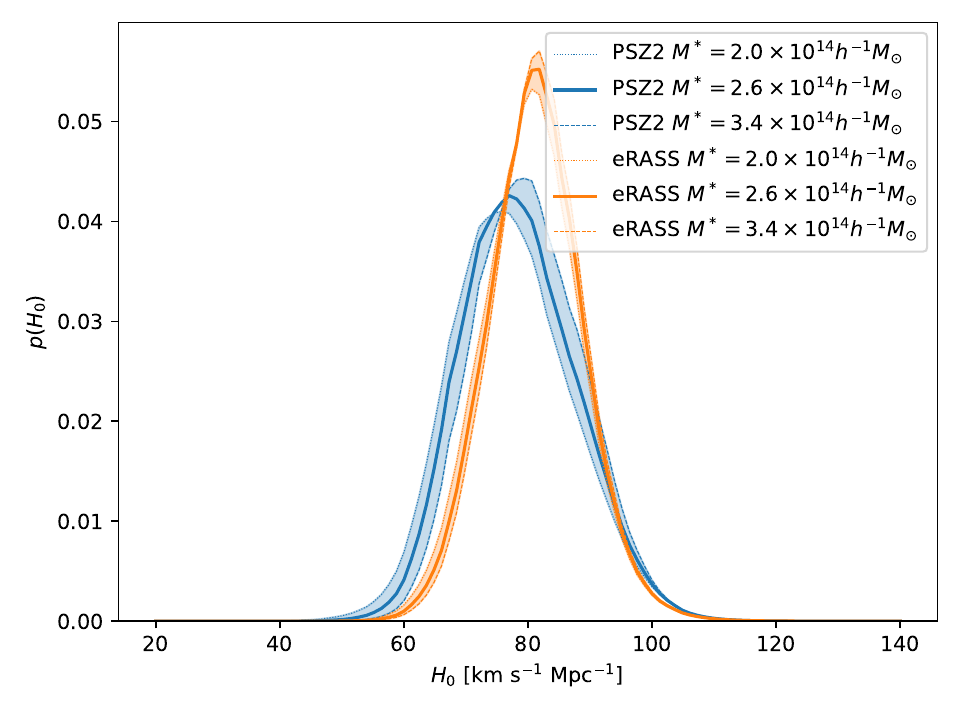}
    \includegraphics[width=0.47\textwidth]{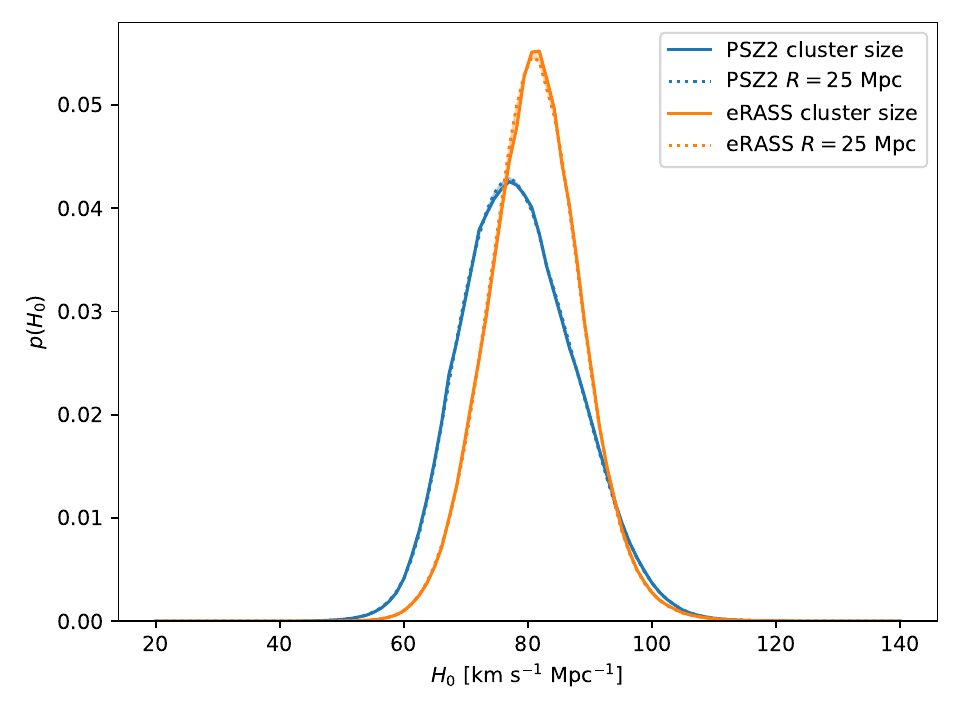}
    \caption{Sensitivity of the PSZ2 and eRASS results to the $M^*$ parameter (top panel) and galaxy cluster redshift uncertainty (bottom panel). We vary the $M^*$ parameter to the limits of its 1-$\sigma$ confidence region of $M^*=2.6^{+0.8}_{-0.6} \times
10^{14} \, h^{-1}M_\odot$. The results are shown as shaded bands in the top panel. Next we increase all the galaxy cluster redshift uncertainties to and $25\,\text{Mpc}$, corresponding to half of typical inter-cluster separations. The results are shown as dotted lines in the bottom panel. No significant difference is observed compared to when the redshift uncertainties from the galaxy clusters are used.}
    \label{fig: systematics}
\end{figure}

In order to provide a preliminary quantification of the associated uncertainties, we carry out two studies. Firstly, we vary the $M^*$ parameter to the limits of its 1-$\sigma$ confidence region of $M^*=2.6^{+0.8}_{-0.6} \times
10^{14} \, h^{-1}M_\odot$. The results are shown as shaded bands in Fig.~\ref{fig: systematics} (top panel). Secondly, we modify the galaxy cluster redshift uncertainties from their values recorded in the corresponding catalogue to a fiducial value of $25\,\text{Mpc}$, i.e., $\Delta z \sim 0.008$, corresponding to half of typical inter-cluster separations. This would mean that a GW event in a void and not associated with a galaxy cluster, is simply assigned to the nearest galaxy cluster. No significant difference in the final results are seen at this stage (Fig.~\ref{fig: systematics} bottom panel).

\begin{figure}
    \centering
    \includegraphics[width=0.47\textwidth]{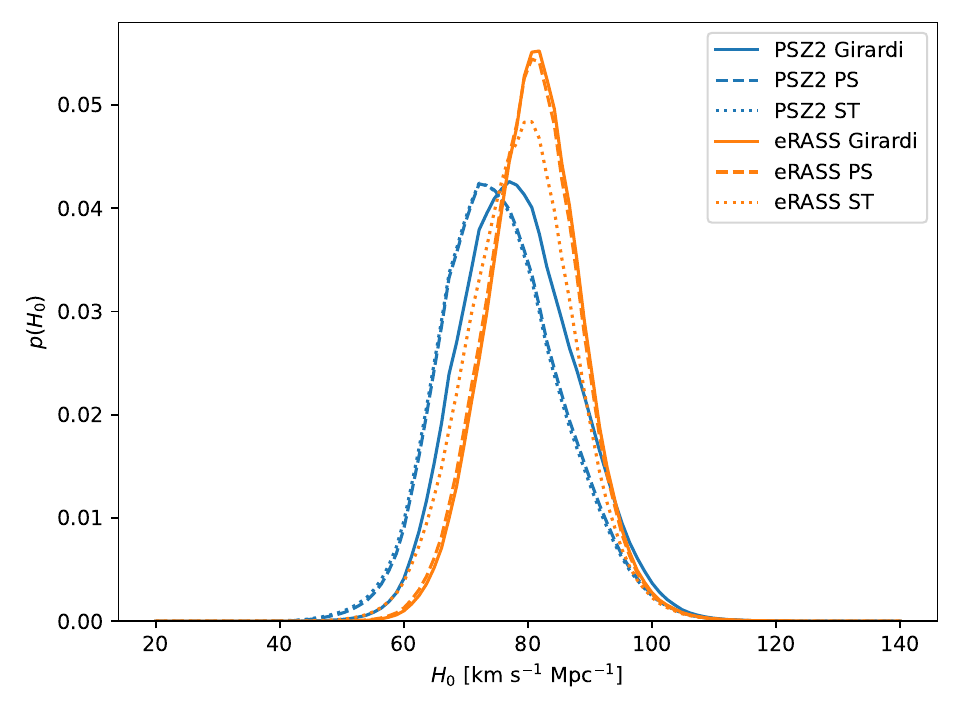}
    \caption{Sensitivity of the PSZ2 and eRASS results to the choice of the cluster mass function, for which we have used the parameterizations in eqs. \eqref{eq: PS}, \eqref{eq: ST} and \eqref{eq: Girardi}, showing the mass function can lead to noticeable shifts of the posterior distributions}
    \label{fig: h0 mass functions}
\end{figure}

The result is also affected by the choice of mass function. This effect is examined by repeating the analysis with the PS and ST mass functions with $n=1$ corresponding to the scale-invariant primordial power spectrum, where $M^*$ has been fitted to the high mass-tails of the cluster mass distributions, yielding $11\times10^{14}\ M_\odot$ and $17\times10^{14}\ M_\odot$ for the PSZ2 and eRASS catalogues respectively. The results are shown in Fig.~\ref{fig: h0 mass functions}, illustrating the noticeable shifts for different mass functions.

Needless to say, this discussion prompts a thorough study of the assumptions and the underlying astrophysics, which is beyond the scope of this initial exploratory work. In particular, the effect of the choice of the mass weighting needs to be investigated further. Moreover one needs to investigate via a set of simulations what the quantitative impact is of GW events not associated with galaxy clusters. On the GW inference side, in particular, in addition to the in-catalogue and out-of-catalogue terms (Sec.~\ref{sec: method}), a third term can be implemented in the LOS-$z$ prior accounting for GW signals originating from galaxies outside of galaxy clusters.

\section{Conclusion and Outlook}
\label{sec: conclusion}

In this work, we show the first results for cosmological inference using only dark siren GW events and EM data from galaxy cluster catalogues instead of galaxies. For the first time, we see significant information coming from the EM sector (beyond mass distributions) for dark standard sirens.
We observe that galaxy cluster catalogues have the potential to improve the $H_0$ measurement precision by reducing the statistical uncertainties by $\sim10$-$38\%$ (see Sec.~\ref{sec: results}). With this extra information we can gain insight into the Hubble tension more quickly in the near-term with upcoming LVK observing runs, as well as probe non-linear cosmology further down the line and into the 3G era of the Cosmic Explorer (CE)~\citep{Evans:2021gyd} and the Einstein Telescope (ET)~\citep{2025arXiv250312263A} as well as with LISA observations~\citep{LISA:2024hlh}. 

CE and ET will observe stellar mass BBHs far beyond $z\gtrsim3$ (potentially out to $z\simeq100$ if such sources are present) and LISA will observe GWs from extreme mass-ratio inspirals (EMRIs) out to $z\simeq 3$, with no EM counterparts expected. \cite{Laghi:2021pqk} showed that with a galaxy catalogue complete at $z=1$, $H_0$ can be estimated with a few percent uncertainty using EMRIs; other cosmological parameters, such as $\Omega_M$ or the dark energy equation-of-state parameter, $w_0$ can also be constrained. Current galaxy catalogues contain almost no information at these redshifts and even upcoming galaxy surveys will not substantially cover this redshift range. Hence it is crucial to explore alternative tracers such as galaxy cluster catalogues for cosmography with GW observations from future detectors. 

A more in-depth study of systematic uncertainties in our analysis needs to be performed (see Sec.~\ref{sec: discussion}). With precise redshifts, direct measurement of masses, and cleaner astrophysical assumptions, galaxy clusters ab initio have the potential to reduce some of the systematic uncertainties in the current method as well and hence improving the $H_0$ measurement accuracy.


Future work towards making this method a viable data analysis pipeline must include a mechanism to include the contribution of a fraction of GW mergers {\em not} hosted in galaxy clusters and a mechanism to construct a LOS-$z$ prior containing information from both galaxy and galaxy cluster catalogues. This will make the formalism applicable and informative for closeby as well as far away GW events. This paradigm can be extended to other galaxy cluster catalogues such as ones obtained by HI intensity mapping, luminous red galaxies (LRGs), quasars, and other alternative redshift tracers. Finally, to fully build confidence in the method, a full set of end-to-end simulations or a mock data challenge (MDC) will be required.

\section*{Acknowledgements}

We would like to thank Martin Hendry for a careful review of the manuscript and Maciej Bilicki, Vasco Gennari, Rachel Gray, Simone Mastrogiovanni, Surhud More, Nicola Tamanini, Cezary Turski, and other members of the LVK CBC Cosmology working group for useful thoughts and comments throughout the course of this study. AG would additionally like to thank Marica Branchesi and Aseem Paranjape for some early discussions on the subject. We would like to thank the organizers of the ICGC 2023 conference where some parts of this project were conceived and the GWCats consortium for several meetings where this work and related aspects were discussed.

The research of FB and AG is supported by Ghent University Special Research Funds (BOF) project BOF/STA/202009/040, the inter-university iBOF project BOF20/IBF/124, and the Fonds Wetenschappelĳk Onderzoek (FWO) research project G0A5E24N. They also acknowledge support to Fonds Wetenschappelijk Onderzoek (FWO) International Research Infrastructure (IRI) grant {\em ``Essential Technologies for the Einstein Telescope''} (I002123N) for funding related to LVK membership and travel. GD acknowledges support from a European Cooperation in Science and Technology (COST) {\em ``CosmoVerse''} short-term scientific mission (STSM) grant which funded a research stay benefitting this project.

This material is based upon work supported by NSF's LIGO Laboratory which is a major facility fully funded by the National Science Foundation in the U.S.A.~and Virgo supported by the European Gravitational Observatory (EGO) and its member states. The authors are grateful for computational resources provided by the LIGO Laboratory and supported by National Science Foundation Grants PHY-0757058 and PHY-0823459.

\appendix

\section{Cosmology, rate and population models}
\label{app: models}

Under the assumptions of homogeneity and isotropy, the luminosity distance can be computed based on the Friedmann-Lema\^itre-Robertson-Walker metric as
\begin{equation}
    d_{L}=\frac{c(1+z)}{H_0} \int_0^z \frac{dz'}{E(z')},
    \label{eq:dlz}
\end{equation}
where $H(z)=H_0 E(z)$ is the redshift-dependent Hubble parameter, and the dimensionless expansion rate $E(z)$ depends on the cosmological model and can be computed for models that obey GR. In this paper, we will restrict our focus to the flat $\Lambda$CDM models, where $E(z)$ is given by
\begin{equation}
E(z)=\left[\Omega_{M}(1+z)^3+\Omega_{{\Lambda}}\right]^{1/2}\,.
\end{equation}
Here, $\Omega_{M}$ is the matter density parameter corresponding to the density in the matter component today relative to the critical density of the Universe. In such models, the dark energy density parameter $\Omega_{\Lambda}=1-\Omega_M$, and the dark energy density are independent of redshift. In models where the theory of gravity is GR, the luminosity distances derived from GW events, $d_{L}^{\rm GW}$, and those based on EM observations, $d_{L}^{\rm EM}$, are identical and given by Eq.~\ref{eq:dlz}.

\begin{table}
\centering
\caption{Summary of the cosmological parameters (fixed in our analysis).}
\label{tab: cosmological params}

\begin{tabular}{cll}
\multicolumn{3}{c}{} \\
\textbf{Parameter} & \textbf{Description} & \textbf{Value} \\
\hline
$\Omega_M$ & Present-day matter density of the Universe&0.3065\\
$w_0$ & Dark energy equation-of-state parameter & -1 \\
\end{tabular}
\end{table}

We construct CBC rate models from independent redshift and source mass distributions as
\begin{equation}
    \begin{aligned} p_{\text {pop }}\left(\theta | \Lambda\right) \propto& p\left(m_1, m_2 | \Lambda_m\right) \\ & \times R\left(z | \gamma, k, z_{\mathrm{p}}\right) \frac{p\left(z | H_0, \Omega_M, w_0 \right)}{1+z},\end{aligned}
\end{equation}
where $p\left(m_1, m_2 | \Lambda_m\right)$ represents the source frame mass distribution, and $\Lambda_m$ includes all parameters describing the mass distribution. The $(1+z)$ term converts the rates between the source and detector frame, and $p(z | H_0, \Omega_M)$ is the redshift prior which is taken to be uniform in comoving volume for the empty catalog case. Here, $\Lambda = \{ \Lambda_m, \gamma, k, z_p, H_0, \Omega_M, w_0\}$ is the set of all  population-level parameters.

The $R(z | \gamma, k, z_{\mathrm{p}})$ term describes the CBC merger rate evolution with redshift with a parametrization similar to \cite{rate_model} which is characterized by a power law slope $\gamma$ and $k$ respectively before and after the peak position denoted by $z_{\mathrm{p}}$.
\begin{equation}
    R \left(z | \gamma, k, z_{\mathrm{p}}\right) = R_0(1+z)^\gamma
    \frac{1+\left(1+z_{\mathrm{p}}\right)^{-(\gamma+k)}}{1+\left[(1+z) /\left(1+z_{\mathrm{p}}\right)\right]^{\gamma+k}}
\end{equation}

\begin{table}
\begin{center}
\caption{Summary of the hyper-parameters used in the merger rate evolution model. The overall merger rate is varied with a $1/R_0$ prior. The other parameters are fixed in our analysis.}
\label{tab: rate model params}
\begin{tabular}{cll}

\textbf{Parameter} & \textbf{Description} & \textbf{Prior/Value} \\
\hline
$R_0$ & Merger rate today in $\mathrm{Gpc}^{-3} \mathrm{yr}^{-1}$ & $1/R_0$\\
$\gamma$ & Slope of power law before the point $z_{\mathrm{p}}$ &  4.59\\
$k$ & Slope of power law after the point $z_{\mathrm{p}}$ & 2.86 \\
$z_{\mathrm{p}}$ & Redshift turning point between power laws & 2.47\\ 
\end{tabular}
\end{center}
\end{table}

The population models are factorized into
\begin{equation}
    p(m_{1,s},m_{2,s}|\Lambda_m)=p(m_{1,s}|\Lambda_m)p(m_{2,s}|m_{1,s},\Lambda_m),
    \label{massprior0}
\end{equation}
where $p\left(m_{1, s} | \Lambda_m\right)$ is the distribution of the primary mass component while $p\left(m_{2, s} | m_{1, s}, \Lambda_m\right)$ is the distribution of the secondary mass component given the primary.

The primary mass is described by a \textsc{Power Law + Peak} model, while the secondary mass component $m_{2, s}$ is described by a truncated power law. 

In our analysis, we fix cosmological parameters (other than $H_0$), the compact binary merger rate evolution, and the compact binary population to values indicated respectively in tables \ref{tab: cosmological params}, \ref{tab: rate model params} and \ref{tab: mass model parameters}. With the latest version of the \texttt{gwcosmo} codebase~\citep{Gray:2023wgj}, one can easily generalize our analysis to situations where the other cosmological as well as the compact binary rate-evolution and population parameters can be varied or marginalized over. The importance of marginalization over the compact binary population and rate evolution parameters has been stressed e.g.~in \cite{Mastrogiovanni:2021wsd,Mastrogiovanni:2023emh,Mastrogiovanni:2023zbw,Gray:2023wgj,Pierra:2023deu}. Since the primary objective of this work is to demonstrate the efficacy of galaxy cluster catalogues, we have chosen to perform a mere exploratory comparison-study, using fixed values for these parameters, as was done in~\cite{LIGOScientific:2021aug}.

\begin{table}
\centering
\caption{Summary of the hyper-parameters in the used \textsc{Power Law + Peak} model. These parameters are fixed in our analysis.}
\label{tab: mass prior params}
\begin{tabular}{cll}
\multicolumn{3}{c}{} \\
\textbf{Parameter} & \textbf{Description} & \textbf{Value} \\
\hline
$\alpha$ &  Power law index primary mass & 3.78\\ 
$\beta$ & Power law index secondary mass & 0.81\\ 
$m_{\rm min}$& Minimum source mass & 4.98\ $M_\odot$\\ 
$m_{\rm max}$&Maximum source mass & 112.5\ $M_\odot$\\ 
$\delta_{\rm m}$& Smoothing parameter &4.8\ $M_\odot$\\ 
$\mu_g$& Peak of Gaussian  &32.27\ $M_\odot$\\ 
$\sigma_g$& Std.~dev.~of Gaussian  &3.88\ $M_\odot$\\ 
$\lambda_{\rm g}$& Fraction of events in Gaussian $\in [0,1]$&0.03\\
\multicolumn{3}{c}{} \\
\end{tabular}
\label{tab: mass model parameters}
\end{table}

\section*{Data availability} 
All of the GW events that are used in this analysis are available at \url{https://gwosc.org/GWTC-3/}. The GLADE+ catalogue is available at \url{https://glade.elte.hu/}. The PSZ2 catalogue is available at \url{https://cdsarc.cds.unistra.fr/viz-bin/Cat?J/A+A/594/A27}. The eRASS catalogue is available at \url{https://cdsarc.cds.unistra.fr/viz-bin/cat/J/A+A/685/A106}.

\bibliographystyle{mnras}
\bibliography{sources}

\bsp
\label{lastpage}
\end{document}